\documentclass[10pt,twocolumn,
superscriptaddress,aip,author-year]{revtex4-1}
\usepackage{times}
\usepackage{color}
\usepackage{amsmath}
\usepackage{amssymb}
\usepackage{graphicx}
\renewcommand{\cite}[1]{\citep{#1}}

\begin{document}

\title{\large Efficient Exploration of the Space of Reconciled Gene Trees}

\author{Gergely J. Sz\"oll\H{o}si}
\email[]{ssolo@elte.hu}
\affiliation{Laboratoire de Biom\'etrie et Biologie Evolutive, Centre National de la Recherche Scientifique, Unit\'e Mixte de Recherche 5558, Universit\'e Lyon 1, F-69622 Villeurbanne, France;}
\affiliation{Universit\'e de Lyon, F-69000 Lyon, France;}
\affiliation{ELTE-MTA ``Lend\"ulet'' Biophysics Research Group
1117 Bp., P\'azm\'any P. stny. 1A., Budapest, Hungary;}
\author{Wojciech Rosikiewicz}
\affiliation{ Laboratory of Bioinformatics, Faculty of Biology, Adam Mickiewicz University, Pozna\'n, Poland}
\author{Bastien Boussau}
\affiliation{Universit\'e Claude Bernard Lyon 1 UMR CNRS 5558 - LBBE, Lyon, France.}
\affiliation{Universit\'e de Lyon, F-69000 Lyon, France;}
\affiliation{Department of Integrative Biology, University of California, Berkeley, California, United States of America.}
\author{Eric Tannier}
\affiliation{Laboratoire de Biom\'etrie et Biologie Evolutive, Centre National de la Recherche Scientifique, Unit\'e Mixte de Recherche 5558, Universit\'e Lyon 1, F-69622 Villeurbanne, France;}
\affiliation{Universit\'e de Lyon, F-69000 Lyon, France;}
\affiliation{Institut National de Recherche en Informatique et en Automatique Rh\^one-Alpes, F-38334 Montbonnot, France}
\author{Vincent Daubin}
\affiliation{Laboratoire de Biom\'etrie et Biologie Evolutive, Centre National de la Recherche Scientifique, Unit\'e Mixte de Recherche 5558, Universit\'e Lyon 1, F-69622 Villeurbanne, France;}
\affiliation{Universit\'e de Lyon, F-69000 Lyon, France;}

\begin{abstract}
Gene trees record the combination of gene level events, such as duplication, transfer and loss, and species level events, such as speciation and extinction. Gene tree-species tree reconciliation methods model these processes by drawing gene trees into the species tree using a series of gene and species level events. The reconstruction of gene trees based on sequence alone almost always involves choosing between statistically equivalent or weakly distinguishable relationships that could be much better resolved based on a putative species tree. To exploit this potential for accurate reconstruction of gene trees the space of reconciled gene trees must be explored according to a joint model of sequence evolution and gene tree-species tree reconciliation.

Here we present amalgamated likelihood estimation (ALE), a probabilistic approach to exhaustively explore all reconciled gene trees that can be amalgamated as a combination of clades observed in a sample of gene trees. We implement the ALE approach in the context of a reconciliation model \cite{szollosi_LGTftD}, which allows for the duplication, transfer and loss of genes. We use ALE to efficiently approximate the sum of the joint likelihood over amalgamations and to find the reconciled gene tree that maximizes the joint likelihood among all such trees.   
     
We demonstrate using simulations that gene trees reconstructed using the joint likelihood are {\color{black} substantially} more accurate than those reconstructed using sequence alone. Using realistic gene tree topologies, branch lengths and alignment sizes, we demonstrate that ALE produces {\color{black} more accurate} gene trees even if the model of sequence evolution is greatly simplified. Finally, examining 1099 gene families from 36 cyanobacterial genomes we find that joint likelihood-based inference results in a striking reduction in apparent phylogenetic discord, with resp. $24\%$,$59\%$ and $46\%$ percent reductions in the mean numbers of duplications, transfers and losses per gene family.  

The open source implementation of ALE is available from \\https://github.com/ssolo/ALE.git .  

\end{abstract}

\maketitle

  \begin{figure*}
  \begin{center}
 \centerline{\includegraphics[width=2.\columnwidth]{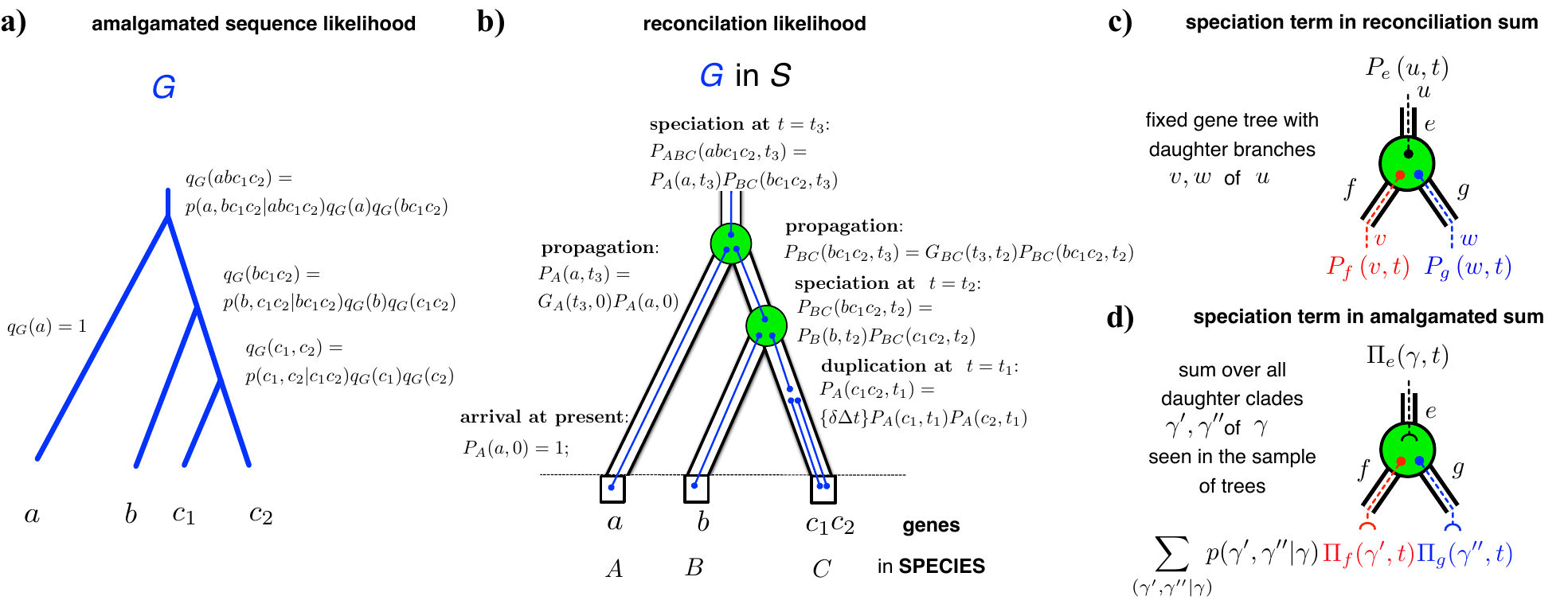}}
  \caption{{\bf Estimating the joint likelihood using amalgamation.} a) based on a sample of gene trees conditional clade probabilities can be used to estimate the posterior probability of a gene tree $G$ that can be amalgamated from clades present in the sample (some terms are not shown).  b) an evolutionary scenario reconciling $G$ with the species tree $S$ that involves a duplication and two speciations. The probability of a scenario, here the probability $P_{ABCD}(abc_1c_2,t_3)$ of seeing the root of $G$ at the root of $S$, can be calculated by using reconciliation events to draw $G$ into $S$ (some terms are not shown). In general we do not know the evolutionary scenario and must sum over  all possible ways to draw $G$ into $S$ to calculate the reconciliation likelihood \cite{szollosi_LGTftD}.  c) the sum over reconciliations is carried out recursively using a set of reconciliation events. Here, we show one such event, a speciation, together with the corresponding term in the probability $P_e(u,t)$ of seeing gene tree branch $u$ in branch $e$ of $S$ at time $t$. d) to extend the recursion to sum over trees that can be amalgamated we have to replace $u$ by the corresponding clade $\gamma$ and sum over all pairs of complementary subclades $\gamma'$, $\gamma''$ present in the gene tree sample.      \label{ALE_scheme}}  
  \end{center}
  \end{figure*} 

Each homologous gene family has its own unique story, but all of these stories are related by a shared species history \cite{Maddison:1997ly,Szollosi:2012fk}. Consequently, knowledge of the pattern of speciations that lead to the species we observe today, i.e., of the species tree, is valuable in gene tree inference. This is the case because sequence data alone often lack enough information to confidently support one gene tree topology over many competing alternatives \cite{Wu:2013fk,Nguyen:2012uq}.   

The problem of how to obtain the species tree itself raises a circular problem: the reconstruction of the species tree requires identifying events of gene family evolution, such as duplications, transfers and losses, and both the reconstruction of gene trees and the identification of such events requires a known species tree. A solution to this problem is a joint inference of gene and species trees, where gene trees reconstructed using candidate species trees are used to infer the species tree itself \cite{Boussau:2010cr,Boussau:2012uq}. Given the plethora of sequence information available, a central element of such an approach is an efficient method capable of reconstructing gene trees given a putative species tree. 

{\color{black}  Here, we present such a method to reconstruct gene trees, which we call amalgamated likelihood estimation (ALE). } The ALE approach allows the combination of the estimation  of sequence likelihood by conditional clade probabilities based on a sample of gene trees \cite{Hohna:2012qf}, with probabilistic reconciliation methods that assume the evolution of gene lineages to be independent \cite{Akerborg:2008uq,tofigh_2009,Rasmussen:2012fk,Szollosi:2012fkkk,Boussau:2012uq,szollosi_LGTftD}. 
We implement the ALE approach in the context of a reconciliation model that considers duplications, transfers and losses \cite{szollosi_LGTftD} by extending the dynamic programming scheme to iterate over the very large number of  reconciled gene trees whose topologies can be amalgamated as a combination of clades observed in the gene tree sample \cite{David:2011zr}.  

{\color{black} To validate our approach we simulate a large number of sequences using gene tree topologies, branch lengths and alignment sizes based on homologous gene families from 36 cyanobacterial genomes. The choice of Cyanobacteria is motivated by i) the availability of a well-resolved \cite{Criscuolo:2011bh}  dated species phylogeny \cite{Szollosi:2012fkkk} and ii) the large evolutionary time spanned by the species tree, with the root dated at $3500-2700$ Mya \cite{Falcon:2010fk}. To perform simulations that are as realistic as possible we use two techniques: First, in a procedure reminiscent of parametric bootstrap methods we infer gene trees using ALE and use these to simulate sequences retaining both alignment sizes and branch length; second, to emulate the complexity of real data, we use a complex model of sequence evolution to simulate sequences, and a simple model to perform reconstructions. }

The simulation results presented below demonstrate that ALE combined with the ODT reconciliation method \cite{szollosi_LGTftD} is able to reconstruct significantly more accurate gene trees compared to reconstruction based on sequence evolution alone. As we show, ALE is more accurate than the sequence-only method even when the latter is run with the correct model of sequence evolution used in the simulations, while ALE relies on a simplified model. Examining reconciliations for the biological dataset on which our simulations are based, we further show that inference using the joint likelihood greatly reduces the number of inferred duplication, transfer and loss events. {\color{black} As we discuss, going beyond the cyanobacterial example, this indicates that the majority of the apparent discord between gene trees may in fact result from uncertainty in reconstructions based on sequence alone.}

\section*{  Materials and Methods  }

\subsection*{Gene Tree Reconciliation using Conditional Clade Probabilities}

Recently, H\"ohna and Drummond \cite{Hohna:2012qf}, and subsequently Larget \cite{Larget:2013fk} demonstrated that conditional clade probabilities (CCP) provide a highly accurate means of approximating posterior probabilities of tree topologies from samples recorded during Markov Chain Monte Carlo (MCMC) sampling. That is, the CCP method accurately approximates the posterior probability of a very large number of gene tree topologies from a converged MCMC run that sampled only a minute fraction of the total tree space. However, it is approximate because, aside of finite sample size, it ignores that the phylogenies of nonoverlapping clades are not necessarily independent of one another. 

The estimation of the posterior probability of a gene tree topology by CCP relies on a  simple recursion during the course of which the tree is incrementally resolved. Consider a rooted bifurcating gene tree $G$. As illustrated in Fig.\ \ref{ALE_scheme}a, for a given clade $\gamma$ the conditional probability $q_G(\gamma)$ of the subtree resolving $\gamma$ in $G$ is    
\begin{equation}
q_G(\gamma) = p(\gamma',\gamma'' | \gamma) q_G(\gamma') q_G(\gamma''),  
\end{equation}
where $\gamma',\gamma'' $ are daughter clades splitting $\gamma$, such that $\gamma \setminus \gamma'  =  \gamma'' $, and $p(\gamma',\gamma'' | \gamma) $ is the probability of observing the split $\gamma'$, $\gamma''$ conditional on $\gamma$ being present. The conditional probability $p(\gamma',\gamma'' | \gamma) $ can be estimated from an MCMC sample as the ratio of the frequency of observing the split implying both daughter clades $f(\gamma',\gamma'')$ and the frequency of observing the mother clade $f(\gamma)$, if clade $\gamma$ is present in the sample, and it is zero otherwise. It follows that $q_G(\gamma)=1$ for clades with a single leaf, which terminate the recursion. The value $q_G(\Gamma)$ for the ubiquitous clade $\Gamma$ comprised of all leaves of $G$ yields the estimate of the posterior probability of $G$. The conditional clade probability is normalised, since summing over all splits $\gamma',\gamma''$ of $\gamma$ at each step of the recursion and $\sum_{(\gamma',\gamma''|\gamma)}   p(\gamma',\gamma'' | \gamma)  =1$ imply $\sum_G q_G(\Gamma) =1$. We refer to gene tree topologies that are comprised of clades observed in an MCMC sample of trees as trees that can be \emph{amalgamated} \cite{David:2011zr}. As defined here, the CCP estimate of the posterior probability is nonzero for trees that can be amalgamated, and zero otherwise. 
           
As illustrated in Fig.\ \ref{ALE_scheme}b,c and d, it is possible to extend probabilistic reconciliation methods that assume the evolution of gene lineages in the species tree to be independent to iterate over the reconciliations of all gene tree topologies that can be amalgamated. Such species tree-gene tree reconciliation methods describe the evolution of a gene family by recursively drawing the corresponding gene tree into the species tree using a series of reconciliation events (Fig.\ \ref{ALE_scheme}c). The reconciliation events used are comprised of one or more atomic event, such as duplication, transfer, loss and speciation, and map branches of the gene tree to branches of the species tree. 

In extending reconciliation methods to consider all possible gene trees that can be amalgamated we are interested in reconciliation events that cause a bifurcation in $G$. Each of these events corresponds to a gene tree branch $u$ being succeeded by its descendants $v$ and $w$.  We replace each such reconciliation event by a series of alternative such events, corresponding to alternative resolutions of the clade $\gamma$ corresponding to $u$. That is, we replace each event that leads to $u$ being succeeded by $v$ and $w$ by a series of events leading to the clade $\gamma$ corresponding to $u$ being succeeded by every split $\gamma',\gamma''$ of $\gamma$ that has been observed in the sample of gene trees used to construct the CCP estimate.  

In the Appendix we develop the ALE approach in the context of the dynamic programming algorithm derived in  \cite{szollosi_LGTftD}. Our goal is to calculate the likelihood of alignment $A$ given the species tree $S$ and a model of gene family evolution $\mathcal{M}_\mathrm{rec.}$ as the sum over gene tree topologies of the product of the posterior probability $P(A|G)$ of the alignment given $G$ and the probability  $P(G |S,\mathcal{M}_\mathrm{rec.} )$ of $G$ given $S$ and $\mathcal{M}_\mathrm{rec.}$ 
\begin{align}
\mathcal{L}_{\mathrm{joint}} (A | S,\mathcal{M}_\mathrm{rec.}) \approx \sum_G P ( A |G ) P(G |S,\mathcal{M}_\mathrm{rec.} ) \nonumber\\
\approx\sum_G \left( q_G(\Gamma) \sum_{e,t} P_e(R,t) \right)
\end{align}
where $P_e(u,t)$ is the probability of seeing branch $u$ of $G$ in branch $e$ of $S$ at time $t$, the sum over $e$ and $t$ corresponds to all species tree branch, time pairs in $S$, and $R$ is the root of $G$.
To calculate $\mathcal{L}_{\mathrm{joint}} (A | S,\mathcal{M}_\mathrm{rec.})$ we use the procedure sketched above to extend the dynamic programming algorithm to simultaneously sum over all reconciled gene trees that can be amalgamated (cf.\ the Appendix). As an example of how this is carried out, consider a speciation event in the species tree $S$ that results in two gene lineages in a fixed gene tree $G$. The corresponding term in the probability of observing the gene tree branch $u$ in branch $e$ of the species tree at time $t$ that the speciation occurs (cf.\  Fig.\ \ref{ALE_scheme}b and equation 6 in \cite{szollosi_LGTftD}) is 
\begin{equation}
P_e(u,t)= \cdots + P_f(v,t) P_g(w,t) + \cdots,  
\nonumber
\end{equation}       
where the $f$ and $g$ are daughters of $e$ in $S$, and $v$ and $w$ are descendants of $u$ in $G$. To calculate the sum of the joint sequence-reconciliation likelihood over all reconciled gene trees that can be amalgamated we replace gene tree branch $u$ with the corresponding clade $\gamma$ and sum over all observed splits $\gamma',\gamma''$ of $\gamma$ weighted by the appropriate conditional probabilities:      
\begin{equation}
\Pi_e(\gamma,t)= \cdots + \sum_{(\gamma',\gamma''|\gamma)} p(\gamma',\gamma''|\gamma) \Pi_f(\gamma',t) \Pi_g(\gamma'',t) + \cdots,  
\nonumber
\end{equation}       
where $\Pi_e(\gamma,t)$ is the probability of observing clade $\gamma$ in branch $e$ of $S$ at time $t$. 

Performing the equivalent procedure for all reconciliation events it follows by recursion that the sum of the joint sequence-reconciliation likelihood $\mathcal{L}_{\mathrm{joint}}$ over all trees $G$ %that can be amalgamated 
is calculated as:       
\begin{align}
\mathcal{L}_{\mathrm{joint}} \approx \sum_G \left( q_G(\Gamma) \sum_{e,t} P_e(R,t) \right) = \sum_{e,t}  \Pi_e(\Gamma,t).  
\label{Lj}
\end{align} 
Reconciled gene trees can be sampled by stochastic backtracking along the sum, while replacing addition by taking the maximum it is possible to find the most likely reconciled tree \cite{Szollosi:2012fkkk}. {\color{black} The calculation of the likelihood (eq.\ \ref{Lj})
takes a few seconds for the data considered in the manuscript.}    

\subsection*{Validation based on ``real" gene trees}          

\begin{figure*}
  \begin{center}
 \centerline{\includegraphics[width=2.\columnwidth]{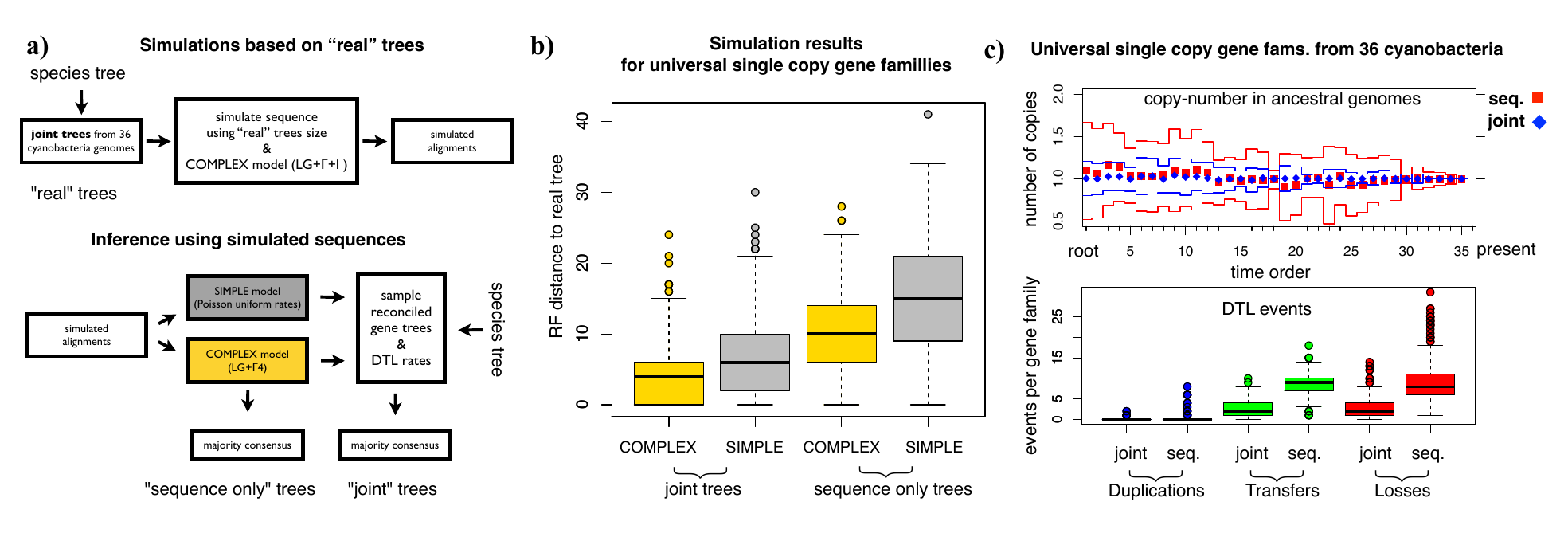}}
  \caption{ {\bf Validating joint likelihood based inference.} a)  we i) reconstructed reconciled gene trees that maximise the joint likelihood using homologous gene families from 36 cyanobacterial genomes together with the species tree from \cite{Szollosi:2012fkkk} show in Fig.\ \ref{S}; ii) simulated sequences using the reconstructed "real" trees and a COMPLEX model of sequence evolution; iii) sampled gene tree topologies using both a SIMPLE model and the COMPLEX model iv) attempted to reconstruct the "real" trees from the simulated sequences using only the sequence alone, and using the joint likelihood together with the species tree for samples from both the SIMPLE and the COMPLEX models. b) computing the Robinson-Foulds distance to the real trees indicates that trees reconstructed from simulated sequences using the joint likelihood are more accurate than those reconstructed based on the sequence alone regardless of the model of sequence evolution used. c) in the top panel we compare the distribution of the number of genes in ancestral genomes based on reconciliations of gene trees reconstructed from $342$ universal single-copy cyanobacterial gene families.  The mean number of copies for joint (blue diamonds) and sequence trees (red squares) is plotted together with the  standard deviation (blue and red lines). The time order of the speciations corresponds to Fig.\ 3 of \cite{Szollosi:2012fkkk}. In the lower panel we compare the number of Duplication, Transfer and Loss events needed to reconcile joint and sequence trees. For details of the inferences presented see Materials and Methods. \label{simfig}}  
  \end{center}
  \end{figure*}

To validate our approach we simulated sequences using tree topologies, branch lengths and alignment sizes based on $1099$ gene families from 36 cyanobacterial genomes available in the HOGENOM database \cite{Penel:2009ly}. As described in detail below and illustrated in Fig.\ \ref{simfig}a, to generate the set of simulated alignments we first reconstructed reconciled gene trees that maximise the joint likelihood and subsequently used the reconstructed gene trees to simulate amino-acid sequences. To emulate the relative complexity of real data compared to available models of sequence evolution we used a complex model of sequence evolution to simulate sequences -- an LG model \cite{Le:2008kx} with across site rate variation and invariant sites, and attempted to reconstruct their history with a simple model  -- a Poisson model \cite{felsenstein81} with no rate variation. 

{\color{black}           
\subsubsection*{Sequence data}

To construct a simulated dataset we first reconstructed gene trees for $1099$ cyanobacterial gene families with $10$ or more genes in any of the $36$ cyanobacteria present in version 5 of the HOGENOM database \cite{Penel:2009ly}. Families with more than $150$ genes were not considered. For each family amino acid sequences were extracted from the database and aligned using MUSCLE(v3.8.31) \cite{Edgar:2004kx} with default parameters. The multiple alignment was subsequently cleaned using GBLOCKS(v0.91b) \cite{Talavera:2007vn} with the options:
\begin{equation}
\text{``{\tt -t=p -b1 50 -b2 50 -b5=a -t=p}''.}\nonumber\label{gblocks}
\end{equation}
Cleaned alignments are available from dryad doi:10.5061/dryad.pv6df.

\subsubsection*{Reconstructing ``real'' trees}
For each cleaned alignment an MCMC sample was obtained using PhyloBayes (v3.2e) \cite{Lartillot:2009uq} using an LG+$\Gamma$4+I substitution model \cite{Le:2008kx}  with a burn-in of $1000$ samples followed by at least $3000$ samples. Following this step gene families were separated into two datasets: i) dataset I, comprised of $342$ universal single-copy families with exactly one copy in each of the $36$ cyanobacteria and, ii) dataset II, which includes dataset I, and is comprised of $1099$ families, each with at least ten genes in any of the $36$ cyanobacterial genomes considered. For the $342$ single-copy universal gene families of dataset I $10000$ trees were sampled

{\color{black}
 For each family we used the species tree shown in Fig.\ \ref{S}, sampled reconciled gene trees using ALEsample (sampling at least $5000$ reconciled trees) to sample DTL rates and reconciled gene trees, and ALEml to find the ML DTL rates and the corresponding ML reconciled gene tree. 
}

For each ALEsample sample we computed the majority consensus tree and fully resolved "real" trees for each gene family were calculated based on the ALEsample sample of trees by finding the tree that maximised conditional clade probabilities based on the sample. For both real and simulated alignments sequence-only trees were also inferred using PhyML (version 20110526) \cite{Guindon:2003ys} using the LG+$\Gamma$4+I model model with the options:    
\begin{equation}
\text{``{\tt -b -4 -m LG -f e -v e -c 4 -a e -s BEST}''.}\nonumber\label{phyml}
\end{equation}

``Real'' gene trees are available from dryad doi:10.5061/dryad.pv6df.
\subsubsection*{Sequence simulation}

To simulate amino acid sequences we used bppseqgen (v1.1.0) \cite{Dutheil:2008fk} keeping the branch lengths and alignment sizes and using the COMPLEX model corresponding to an LG model with site rate variation described by a Gamma-distribution with $\alpha=0.1$ and $10\%$ invariant sites.

Simulated alignments are available from dryad doi:10.5061/dryad.pv6df.

\subsubsection*{Inference for simulated data}

For each simulated alignment an MCMC sample was obtained using PhyloBayes (v3.2e) using  i) a SIMPLE model corresponding to a Poisson model \cite{felsenstein81} with no rate variation. 

We sampled $10000$ trees after a burnin of $1000$  samples with a sample taken every $10$ iterations. For the simulated sequence corresponding to the $342$ single-copy universal gene families of dataset I
we also sampled trees using the COMPLEX model corresponding to an LG+$\Gamma$4+I substitution model, sampling  $3000$ trees after a burnin of $1000$  samples.

For each family we sampled reconciled gene trees using ALEsample (sampling at least $5000$ reconciled trees) to sample DTL rates and reconciled gene trees, and ALEml to find the ML DTL rates and the corresponding ML reconciled gene tree. 

Distances to the ``real'' tree for gene trees of dataset I (Fig.\ \ref{simfig}b) were computed as the distance between majority consensus trees calculated from the sequence-only PhyloBayes samples for both the SIMPLE and the COMPLEX model as well as the joint ALEsample samples for both. The same procedure was used for the simulated sequence corresponding to dataset II (Fig.\ \ref{figA1}a)  for the SIMPLE model. For the COMPLEX model joint trees were not computed and PhyML trees were used for the sequence-only trees.

\subsubsection*{Inference of numbers of DTL events}

The number of DTL events for joint trees was inferred using ALEml using a samples of trees obtained using the SIMPLE model. The number of DTL events for sequence trees was inferred using ALEml using fixed PhyML trees (based on LG+$\Gamma$4+I substitution model).     

ML reconciled trees are available from dryad doi:10.5061/dryad.pv6df.

\subsubsection*{Statistical support}
Statistical support of bipartitions was calculated from samples of gene trees obtained either using PhyloBayes, for the sequence-only case, or using ALEsample in the joint case. The support of each observed bipartition was estimated as the fraction of all trees in which it was present.   
}
\section*{ Results }

\subsection*{Analysis of gene families from 36 Cyanobacteria}          
{\color{black}
As described above we performed simulations based on two datasets: i) dataset I, comprised of $342$ universal single-copy families with exactly one copy in each of the $36$ cyanobacteria and, ii) dataset II, which includes dataset I, and is comprised of $1099$ families, each with at least ten genes in any of the $36$ cyanobacterial genomes considered. As shown in Fig.\ \ref{simfig}b and Fig.\ \ref{figA1}a of the appendix, for both datasets gene tree reconstruction based on the joint likelihood substantially improves accuracy in comparison to inference based on sequence alone. In fact, we found that the joint reconstruction based on the simple model of sequence evolution yielded significantly more accurate gene trees than the sequence-only inference relying on the complex model used to simulate the alignments. 
}

In our inference on biological data, we chose to consider separately the universal single-copy gene families of dataset I because -- since these families have exactly one copy in all extant cyanobacteria -- we can expect that they were also present in a single-copy in ancestral genomes. Testing to what extent this assumption is satisfied allows us to assess the accuracy of gene trees reconstructed from real-life gene sequences, where we do not have knowledge of the correct tree. An equivalent assumption cannot be made for all families in dataset II, i.e., families that are multi-copy families and/or have a more limited distribution in extant species. As show in Fig.\ \ref{simfig}c gene trees reconstructed using joint likelihood imply that the number of gene copies in ancestral genomes is very close to one with, e.g.\ $328$ families with one, only $6$ families with zero gene copies and $8$ with more than one copy at the root. In contrast, for gene trees inferred based on sequence information only, $248$ families have one, $34$ families have zero gene copies and $60$ have more than one copy at the root of Cyanobacteria.

Considered together with the simulation result the reconciliations of universal single-copy families not only demonstrate that ALE is able to reconstruct accurate gene trees, but also suggests that gene trees inferred using the joint likelihood are significantly different from gene trees inferred based on sequence alone. The magnitude of this difference is reflected in the number of duplication, transfer and loss (DTL) events that are required to reconcile the two sets of gene trees with the species tree. In dataset I the reduction in the number of events necessary to reconcile joint trees is $81.6\%$ for duplications, $70.9\%$ for transfers and $70.2\%$ for losses. In dataset II the reduction in the number of required events is $24.3\%$ for duplications, $59.1\%$ for transfers and $45.8\%$ for losses. The validity of these results is supported by simulation results, where we find that the number of duplications and transfers per family for trees inferred using the joint likelihood is accurately recovered. As shown in Fig.\ \ref{figA1}b  the number of duplications and transfers needed to reconcile joint trees is statistically indistinguishable ($p>0.1$ for both paired T and Wilcox sign rank tests) from the corresponding number of events needed to reconcile ``real" trees used to simulate the alignments. The number of losses per tree are slightly less accurately recovered with an \emph{increase} of $12.1\%$ in the number of events needed to reconcile joint trees.       

Consistent with the above result we find that the distance to the species tree is recovered accurately in our simulations. For simulations based on the $342$ single-copy universal families the Robinson-Foulds  distance to the species tree for ``real'' gene trees has a mean of $11.41$, while the corresponding fully resolved maximum likelihood (ML) reconciled gene trees reconstructed based on the SIMPLE sequence evolution model have a moderately \emph{increased} distance to the species tree with a mean of $13.02$, in comparison the mean distance of sequence-only trees reconstructed using the COMPLEX and SIMPLE models are respectively $17.77$ and $21.80$ (cf.\ Fig.\ \ref{simscale}).        

{\color{black}
A possible concern regarding the joint inference is that we may overfit the species tree. As shown in Fig.\ \ref{simscale} in simulations the distance of the reconstructed trees to both the real tree and the species tree exhibits a decreasing trend for increasing sample size, with no sign of overfitting for any sample size. However, based on Fig.\ \ref{simscale} alone we cannot rule out that overfitting of the species tree would not occur for larger sample sizes. A possible test that does not involve a computationally expensive increase in sample size is to examine the correlation between reconstruction accuracy and alignment size. If overfitting is present we expect it to be stronger for shorter alignments. Such a trend is not observed in our data, in fact, for the largest sample size considered alignment length is negatively correlated with reconstruction error, measured as either i) the distance to the real trees (Pearson's $r=-0.44$ with $p<10^{-5}$); or ii) the difference of the distance of the reconstructed tree and the real tree to the species tree (Pearson's $r=-0.20$ with $p< 10^{-3}$). In other words, reconstructions based on shorter simulated alignments are less accurate and are on average more distant from the species tree than real trees. Such an explicit test is only possible for simulated alignments, however we do observe that the distance to the species tree of real trees (reconstructed from cyanobacterial sequences) are not correlated with alignment length (Pearson's $r=-0.0148$ with $p=0.78$). 
}

\subsection*{Analysis of the signal for the phylogenetic discord }   

\begin{figure}
   \begin{center}
     \centerline{\includegraphics[width=1.\columnwidth]{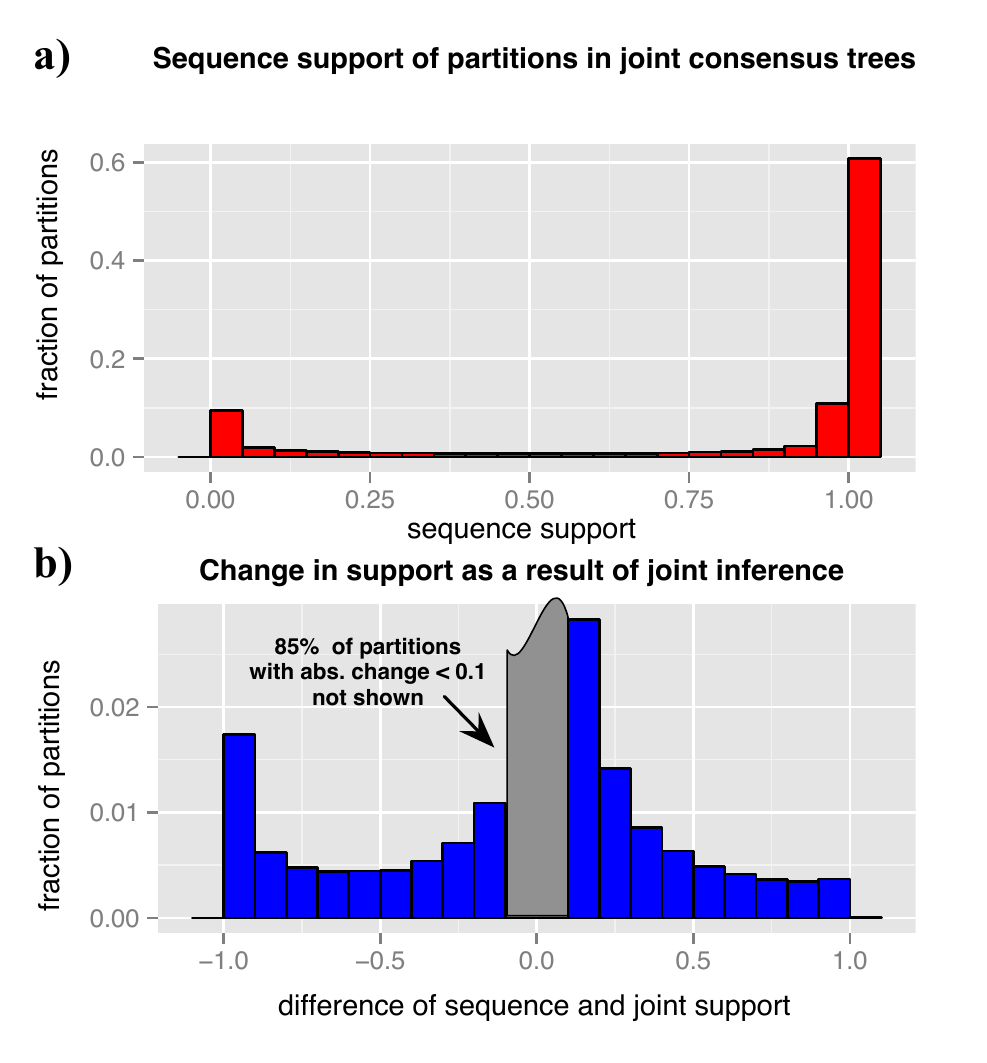}}
     \caption{{\bf Statistical support for 1099 gene trees from 36 cyanobacteria.} We calculated the statistical support of bipartitions as their frequency in MCMC samples based on both the joint likelihood and sequence alone. a) shows the distribution of sequence-only support for bipartitions present in the joint majority consensus trees. b) presents the distribution of the difference between sequence-only and joint support for all bipartitions. \label{sup}}  
   \end{center}
 \end{figure}

{\color{black}
 Considering the above, the results of joint inference present strong evidence that the majority of apparent phylogenetic discord observed among gene trees based on sequence information alone results from reconstruction uncertainty.} To examine the signal for the phylogenetic relationships responsible for the spurious discord we computed the statistical support of bipartitions based on sequence alone as well as based on joint likelihood. As shown in Fig.\ \ref{sup}a most of the bipartitions present in consensus trees based on the joint likelihood are also supported according to the sequence, with $71\%$ of bipartitions in joint trees having a statistical support $>0.95$ according to sequence alone. A significant minority of the bipartitions in joint consensus trees are, however, not supported by the sequence, with $6.4\%$ of bipartitions  in joint trees having a statistical support $>0.95$ according to the joint likelihood, but $<0.05$ according to sequence alone.  Examining the statistical support of partitions in simulations we observe very similar results (cf. Fig.\ \ref{simsup}a). 

To quantify how often the opposite case occurs, i.e., how often are bipartitions strongly supported by sequence rejected based on the joint likelihood we computed the change in statistical likelihood as a result of joint inference. As show in Fig.\ \ref{sup}b the difference of the support according to sequence alone and the support according to the joint likelihood is small for most bipartitions, with $85.8\%$ of bipartitions having an absolute difference $<0.1$. Examining the remaining bipartitions an excess of partitions with a difference $<-0.95$ is present (left corner of Fig.\ \ref{sup}b), comprised of $1.4\%$ of all observed bipartitions. These are partitions that are not supported by sequence, but are strongly supported based on joint likelihood. There is, in contrast, no excess in the number of partitions with a difference $>0.95$  (right corner of Fig.\ \ref{sup}b), corresponding to partitions that are strongly supported by sequence, but are not supported based on joint likelihood, with only  $0.18\%$ of partitions having a difference $>0.95$. Examining the statistical support of partitions in simulations we observe very similar results (cf. Fig.\ \ref{simsup}b). 

\section{Discussion}          

We present a probabilistic method, which we call amalgamated likelihood estimation (ALE), that is able to exhaustively explore the joint likelihood of a very large number of reconciled gene trees using a sample of trees comprising only a minute fraction of the total tree space. We implement ALE in the context of one of the most general gene tree - species tree reconciliation methods available that allows for the duplication, transfer and loss of genes \cite{szollosi_LGTftD}. The general computational scheme, however is applicable to other models considering, e.g.\ duplication and loss \cite{Akerborg:2009fk,Boussau:2012uq}, lineage sorting \cite{Edwards:2007uq,Liu:2007kx} or both \cite{Rasmussen:2012fk}.

To validate our implementation  we simulate sequences based on homologous gene families from $36$ completely sequenced cyanobacterial genomes.
% The choice of Cyanobacteria is motivated by i) the availability of a well-resolved \cite{Criscuolo:2011bh}  dated species phylogeny \cite{Szollosi:2012fkkk} and ii) the relatively large evolutionary time spanned by the species tree, with the root dated at $3500-2700$ Mya \cite{Falcon:2010fk}. To perform simulations that are as realistic as possible we use two techniques: First, in a procedure reminiscent of parametric bootstrap methods we infer gene trees using ALE and use these to simulate sequences retaining both alignment sizes and branch length; second, to emulate the relative complexity of real data, we use a complex model of sequence evolution to simulate sequences, and a simple model to perform reconstructions.
 Contrasting the simulated and the real data sets we find that both the statistical support of simulated and real gene trees (cf.\ Fig.\ \ref{figA1}c and f) and the topological distance between sequence and joint trees are comparable (the mean Robinson-Foulds distance between joint and sequence trees is $13.25$ for simulations and $19.12$ for real data).          

Simulation results together with reconciliations for universal single-copy gene families from 36 cyanobacteria, both presented in Fig.\ \ref{simfig}, establish that ALE reconstructs gene trees that are more accurate than those based on sequence alone. Examining the statistical support for gene trees for both the real and the simulated dataset, we can conclude that overall: i) the majority of relationships inferred from sequence alone are also found in joint trees, with $88.5\%$ of bipartitions ($90.7\%$ in simulations) shared among the two sets of consensus trees, but ii) a significant minority of bipartitions in joint phylogenies have low sequence support, with $9.5\%$ ($7.5\%$ in simulations) having a sequence support $<0.05$, and iii) more rarely, relationships that are strongly supported by sequence are not found in joint consensus trees, with $1.9\%$ of bipartitions  ($1.5\%$ in simulations)  with sequence support $>0.95$ missing from joint trees, and finally iv) joint trees are significantly better supported than sequence trees with $90.3\%$ vs.\ $80.0\%$ of bipartitions in consensus trees ($92.4\%$ vs.\ $83.6\%$ in simulations) having a support $>0.95$. 

There are two intrinsic limitations to the accuracy of ALE-based inferences. First, ALE is approximate in that conditional clade probabilities on which it relies reconstruct the posterior probability of gene trees from marginal frequencies of splits, assuming conditional clade probabilities to be independent. However, while this independence assumption is in general false, H\"ohna and Drummond have demonstrated that in practice CCP estimates based on sufficiently large  samples of trees usually give very accurate approximations of the posterior probabilities \cite{Hohna:2012qf}. Furthermore, as we demonstrate in the appendix, ignoring dependencies between clades is not an arbitrary assumption, but the CCP-based estimate of the posterior probability in fact corresponds to the maximum entropy distribution \cite{Jaynes:2003fk} given marginal split frequencies observed from an MCMC sample. Second, and from a practical point of view more importantly, ALE-based inferences rely on a finite sample of tree topologies, between $3000$ and $10000$ in the results presented here. The corresponding number of amalgamations considered can be very large, e.g.\ for the cyanobacterial gene families considered here up to $10^{40}$, with a geometric mean of $\approx10^{12}$. Despite the large number of amalgamations we find in simulations that only $98\%$ of bipartitions comprising "real" gene trees are present in sampled trees. The correlation between reconstruction error (the distance of the reconstructed tree to the real tree) and the fraction of missing bipartitions is high and significant (Pearson's $r=0.71$ with $p<10^{-5}$). This suggests that the accuracy of ALE-based reconstructions can be significantly further improved by increasing the size and/or diversity of the underlying MCMC samples (also cf.\ Fig.\ \ref{simscale}).  

From the perspective of gene tree-species tree reconciliation we find that, as shown in Fig.\ \ref{simfig}c and  Fig.\ \ref{figA1}e,  joint inference results in a dramatic reduction in the number of events required to describe the evolution of gene trees along the species tree. This decrease is particularly remarkable for the number of transfer events (which make up $69\%$ of the birth events) with only $3.6$ transfers per family in joint trees, compared to $8.7$ for sequence trees in dataset II. The reduction in the number of transfers is reflected in a striking drop in phylogenetic discord, corresponding to an over two-fold reduction in the Robinson-Foulds distance of the species tree and gene trees for single-copy universal families (from $25.8$ to $11.4$, cf.\ Fig.\ \ref{figA1}d).

Obtaining results similar to the above for bacterial or archaeal phyla other than the cyanobacteria is currently limited by the availability of well-supported dated species phylogenies. Joint inference of species and gene trees offers a path toward surmounting this obstacle \cite{Boussau:2010cr,Boussau:2012uq,Szollosi:2012fk,Szollosi:2012fkkk}. {\color{black}
 However, as there is no reason to believe that results for other groups will be qualitatively different, we believe that our results strongly suggest that the majority of apparent phylogenetic discord is the result of uncertainty in phylogenetic reconstructions not only for cyanobacteria, but other groups as well.  

In summary, we find that the majority of phylogenetic discord results from uncertainty in sequence-based reconstruction that can be corrected using information aggregated across gene families by a putative species tree.} Finally, as a corollary of the observation that gene trees reconstructed by combining a simplistic model of sequence evolution with a reconciliation method are more accurate than trees reconstructed using the correct sequence evolution model, we note that while developing increasingly sophisticated models of sequence evolution is of fundamental interest, the potential of probabilistic models of species tree-gene tree reconciliation remain nearly untapped.             
  
%This does not mean that the mechanism whose action is recorded by differences between gene trees, in particular lateral gene transfer, have not played an important role in the evolution of unicellular genomes. The result, however that up to two thirds of transfer events suggested by sequence alone are spurious does demonstrate the important role of probabilistic reconciliation methods and joint inference in establishing just how much gene transfer has occurred, and between which groups.      

\begin{acknowledgments}
We thank Nicolas Lartillot for his suggestions about the MaxEnt derivation. We thank all members of the Bioinformatics and Evolutionary Genomics Group for discussions of the results and comments on the manuscript. GJSz was supported by the Marie Curie Fellowship 253642 ``Geneforest'' and the Albert Szent-Gy\"orgyi Call-Home Researcher Scholarship A1-SZGYA-FOK-13-0005. BB has been supported by a Human Frontier Science Program fellowship. This work was granted access to the Institut National de Physique Nucl\'eaire et de Physique des Particules' (IN2P3) computing centre. This project was supported by the French Agence Nationale de la Recherche (ANR) through Grant ANR-10-BINF-01-01 ``Ancestrome''.
\end{acknowledgments}

\bibliographystyle{sysbio}

\clearpage
\newpage
\appendix
\renewcommand{\thefigure}{A\arabic{figure}}
\addtocounter{figure}{-3}
\renewcommand{\theequation}{A\arabic{equation}}
\addtocounter{equation}{-6}

\section{Appendix}

 \begin{figure*}
  \begin{center}
 \centerline{\includegraphics[width=2.\columnwidth]{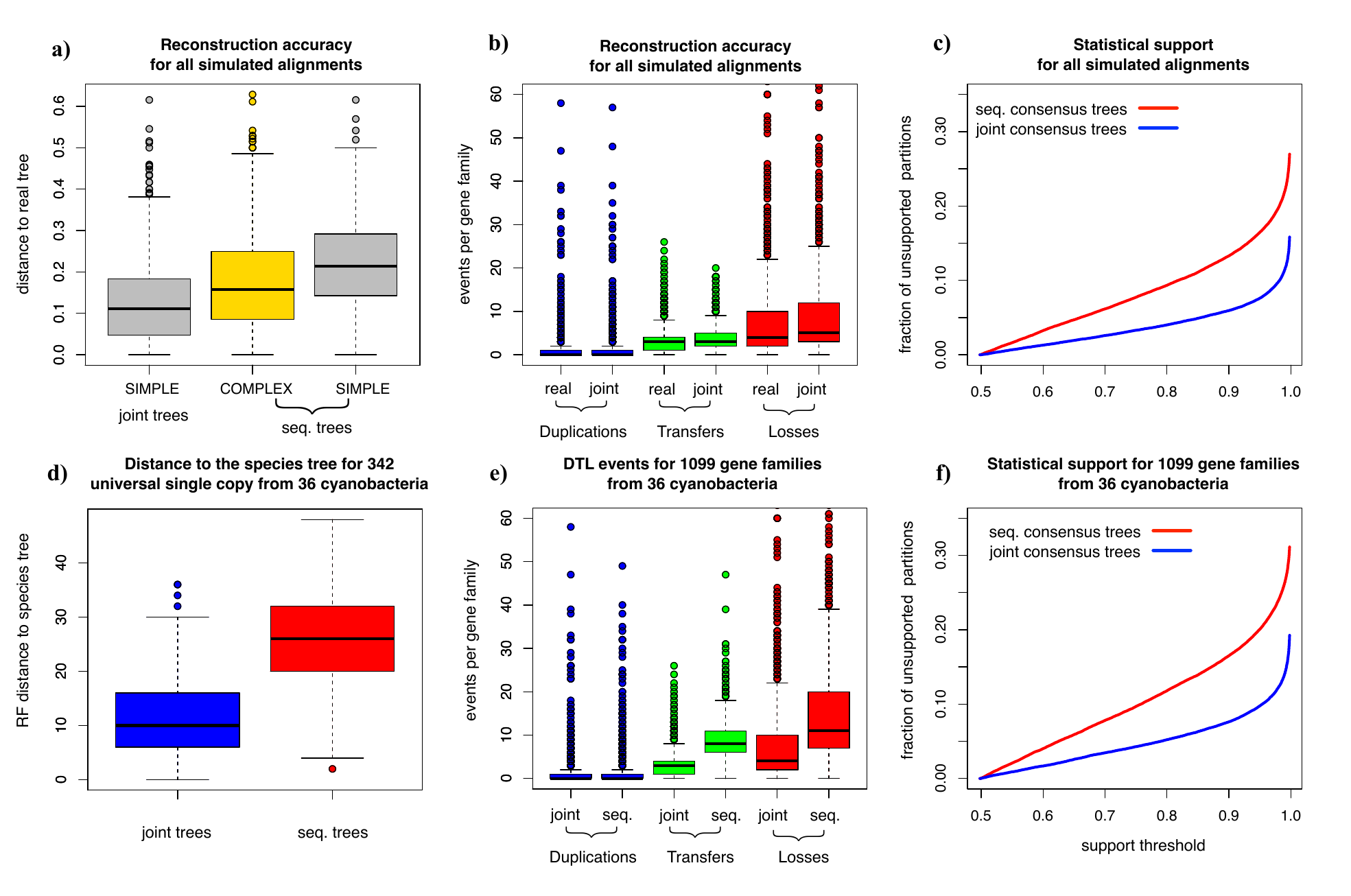}}
  \caption{{\bf Results of joint likelihood-based reconstruction for simulated and real data.} a) the distribution of normalised the Robinson-Foulds distance to the real tree used to simulate sequences, defined as the distance divided by its maximum possible value in each gene tree, for all simulated gene families. Joint inference based on the COMPLEX model was only performed for single-copy universal families (cf.\ Fig.\ \ref{simfig}b). b) comparison of the distribution of DTL events for all simulated gene families. Some points fall outside the range of the ordinate. c) the fraction of bipartitions in majority consensus trees with statistical support over a given threshold for all simulated gene families. d) Robinson-Foulds distance to the species tree for $342$ single-copy universal gene families from $36$ cyanobacterial genomes. e)  DTL events for $1099$ gene families from $36$ cyanobacterial genomes. Some points fall outside the range of the ordinate. f)   the fraction of bipartitions in majority consensus trees with statistical support over a given threshold for $1099$ gene families from $36$ cyanobacterial genomes.\label{figA1}}  
  \end{center}
  \end{figure*} 

 \begin{figure}
   \begin{center}
 \centerline{\includegraphics[width=1.\columnwidth]{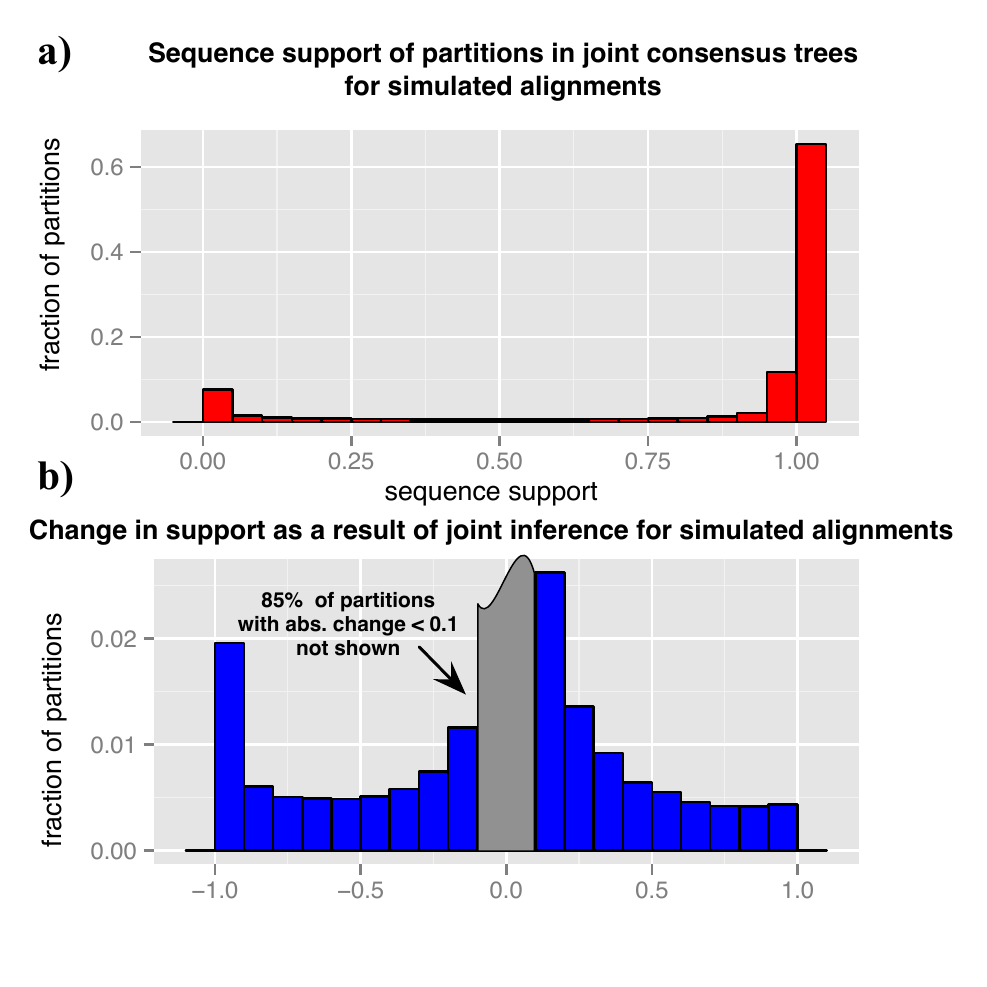}}
 \caption{{\bf Statistical support for simulated gene families.} We calculated the statistical support of bipartitions as their frequency in MCMC samples based on both the joint likelihood and sequence alone. a) shows the distribution of sequence-only support for bipartitions present in the joint majority consensus trees. b) presents the distribution of the difference between sequence-only and joint support for all bipartitions.   \label{simsup}}  
 \end{center}
 \end{figure} 

 \begin{figure}
   \begin{center}
 \centerline{\includegraphics[width=1.\columnwidth]{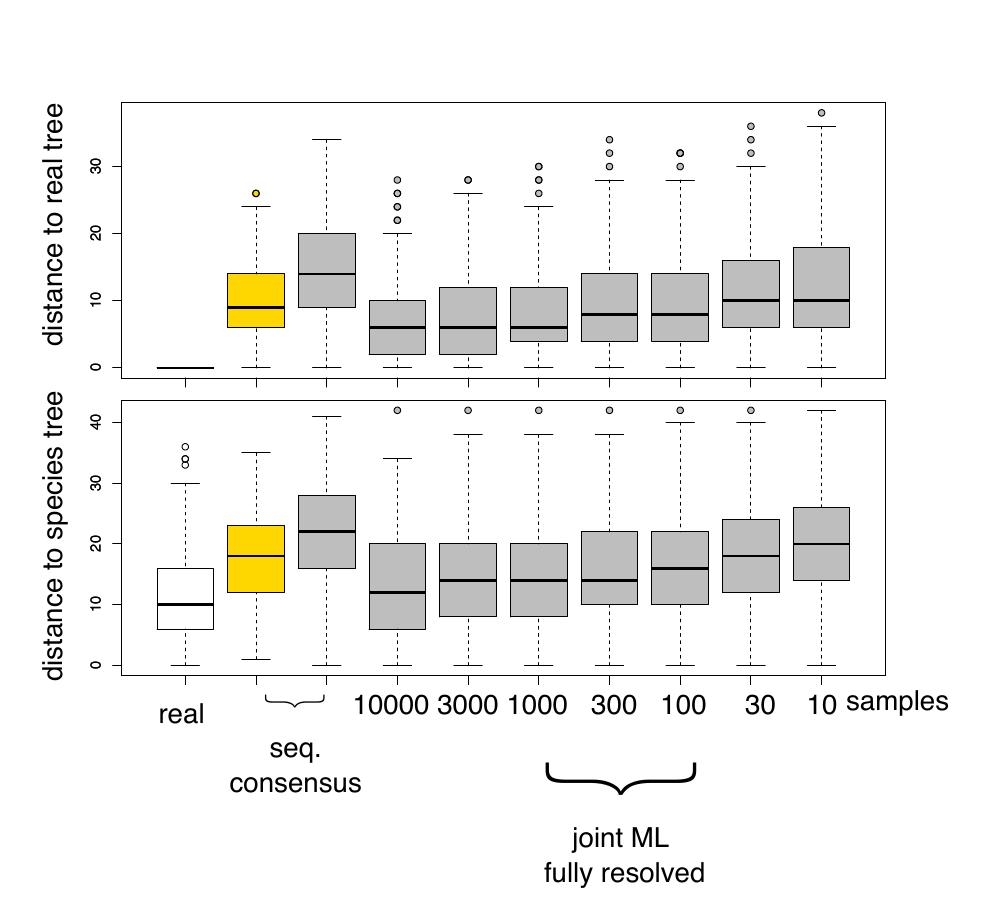}}
 \caption{{\bf Reconstruction accuracy for different sample sizes.} To examine the accuracy of reconstructions for simulated data we used ALEml to recover the ML reconciled trees for 342  universal single-copy families from simulated sequences. In both the top and bottom panel the first set values in white corresponds to real trees. The second and third set of values were obtained from sequence-only samples for respectively the COMPLEX and SIMPLE models of sequence evolution. The seven remaining set of values correspond to ALEml estimates of the ML reconciled trees for samples of $10,30,100,300,1000,3000,10000$ gene tree chosen randomly and without replacement. \label{simscale}}  
 \end{center}
 \end{figure} 

 \begin{figure*}
   \begin{center}
 \centerline{\includegraphics[width=2.\columnwidth]{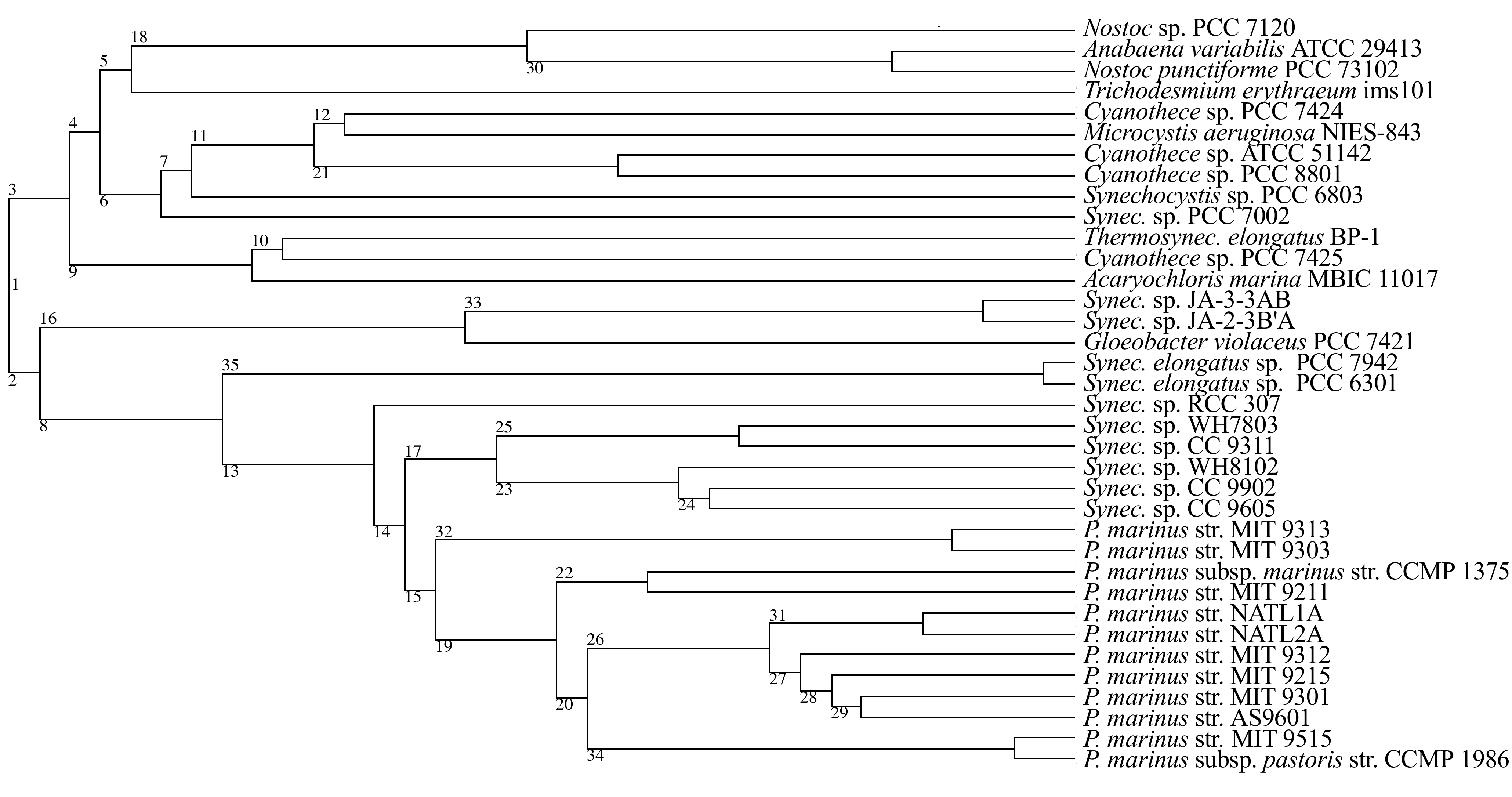}}
 \caption{ {\color{black}
 {\bf Chronologically ordered species tree used in gene tree inference.}  Maximum likelihood chronologically ordered species phylogeny based on 36 genomes with 8332 homologous gene families from \cite{Szollosi:2012fkkk}.\label{S} }}  
 \end{center}
 \end{figure*}

\section*{A minimal model of speciation and gene birth and death}

In the presence of lateral gene transfer (LGT), gene trees record evolutionary paths along the complete species tree, including extinct and unsampled branches, and not only along the phylogeny of the species in which they reside today. This is the case because, while LGT events imply that the donor and receiver lineages existed at the same time, the donor lineage might have subsequently become extinct, or more generally, might not have been sampled. However, it is not feasible to specify, much less to reconstruct, the complete phylogeny of all species that ever existed. Hence, to describe the evolution of genes outside the represented phylogeny -- along lineages that have become extinct or whose descendants have not been sampled -- we must resort to modelling the speciation dynamics that gave rise to the complete phylogeny \cite{szollosi_LGTftD}.  

As a minimal model of speciation, in \cite{szollosi_LGTftD}  we assume that the number of species $N$ is constant, and that the dynamics of speciation are modeled by a continuous time Moran process. That is, for each species at rate $\sigma$, a speciation occurs during which the species gives rise to two descendants and a randomly chosen species goes extinct. We further assume that, of the $N$ species existing at present, we sample only a small fraction $n \ll N$. These $n$ species are the species represented in $S$.

To describe the evolution of genes within the genomes of species we assume genes to evolve independently according to a birth-and-death process that consists of gene duplication, transfer and loss. A gene in the genome of any of the $N$ species can: i) be duplicated at rate $\delta$; ii) be transferred from a donor species to any of the other $N-1$ possible host species at a rate $\tau/(N-1)$; or iii) be lost at a rate $\lambda$. Genes copies can also be born and be lost as a result of the speciation dynamics: iv) at the species level lineages experience speciation at a rate $\sigma$, in which case they are replaced by two copies in the two new species, or v) suffer extinction at an identical rate $\sigma$. A branch $e$ of the represented tree $S$ in general corresponds to a series of speciation events, however, only the last one of these, the speciation event that gave rise to two represented lineages, is explicitly present for internal branches as the speciation node terminating an internal branch of $S$.

\subsection{Amalgamated Sum Over Reconciled Trees}

As developed in \cite{szollosi_LGTftD}, in order to derive the recursion expressing the probability of $G$ as the sum over possible paths along the species tree $S$ we discretize time along $S$ using the series of speciation times $t_i$ along $S$, with $t_0=0$ corresponding to the root of $S$ and $t_n=0$ to the present. Speciations represented in $S$ define the time intervals $[0,t_{1}),\dots,[t_i,t_{i+1}),\dots [t_{n-1},t_{n-1})$ referred to as \emph{time slices} with indices $0,\dots, i, \dots n$. We further divide each time slice into $D$ equal time intervals of height $\Delta t_i = (t_{i+1}-t_i) / D $. 

First, we must describe the evolution of gene copies that appear as single gene lineages when observed from the present. We have to calculate: i) the extinction probability $E_e(t)$ that a gene lineage seen at time $t$ on branch $e$ of $S$ leaves no \emph{observed descendant} i.e., no descendant exists at time $t=0$ in the genome of any of the $n$ sampled species; ii) the extinction probability $\bar E(t)$ that a gene seen at time $t$ in an unrepresented species leaves no observed descendant; iii) the single gene propagation probabilities $G_e(s,t)$ that all observed descendants of a gene seen at time $s$ on branch $e$ descend from a descendant seen at a later time $t<s$ on branch $e$; and iv) $\bar G(s,t)$ the probability that all observed descendants of a gene seen at time $s$ in an unrepresented species descend from a descendant seen at time $t<s$ in an unrepresented species. Differential equations that can be used to calculate the above functions are available in the Appendix of \cite{szollosi_LGTftD}.

Using the extinction probabilities and single gene propagators we sum over all reconciled trees that can be amalgamated by recursively mapping the branches of $G$ onto branches of $S$, as well as unrepresented species using the set of reconciliation events from \cite{szollosi_LGTftD}. 

The probability of the lineage leading to the first bifurcation resolving clade $\gamma$ being seen on branch $e$ of $S$ at time $t_i+\Delta t_i$ given the probabilities at time $t_i$ is 
\begin{align}
&\Pi_e (\gamma,t_i+\Delta t_i) = G_e (t_i+\Delta t_i,t_i ) \Pi_e(\gamma,t_i) \label{Pie}\\
& +\left\{\delta \Delta t_i \right\} \sum_{(\gamma',\gamma''|\gamma)} p (\gamma',\gamma''|\gamma)\Pi_e(\gamma',t_i) \Pi_e(\gamma'',t_i) \nonumber\\
& +\left\{\sigma \Delta t_i \right\} \sum_{(\gamma',\gamma''|\gamma)} p (\gamma',\gamma''|\gamma) \bar \Pi(\gamma',t_i) \Pi_e(\gamma'',t_i)\nonumber \\
& +\left\{\sigma \Delta t_i \right\} \sum_{(\gamma',\gamma''|\gamma)} p (\gamma',\gamma''|\gamma)\Pi_e(\gamma',t_i) \bar \Pi(\gamma'',t_i)\nonumber \\
& +\left\{\sigma \Delta t_i \right\} \bar \Pi(\gamma,t) E_e(t_i), \nonumber 
\end{align}
where $\bar \Pi(\gamma,t)$ denotes the probability of the gene lineage leading to the first bifurcation resolving clade $\gamma$ being seen in an unrepresented species at time $t$, and the sum goes over all splits $\gamma'$,$\gamma''$ of $\gamma$ observed in the MCMC sample used to construct the CCP estimate. The terms correspond to i) no event with an observed descendent; ii) birth of two gene lineages by duplication, such that both leave observed descendants; iii) and iv) birth of two gene lineages with observed descendants as a result of an unrepresented speciation; and finally, v) unrepresented speciation followed by the loss of the copy in branch $e$ such that only the copy in the unrepresented phylogeny leaves an observed descendant, cf.\ Eq.4 and  Fig.\ A1 of \cite{szollosi_LGTftD}.          

The probability of the lineage leading to the first bifurcation resolving clade $\gamma$ being seen in an unrepresented species is:
\begin{align}
& \bar \Pi(\gamma,t_i+\Delta t_i)  = \bar G (t+ \Delta t_i,t_i ) \bar \Pi(\gamma,t_i) \label{Pia}\\
& + \left\{ (2 \sigma + \delta + \frac{N-n_i }{N-1}\tau )\Delta t_i \right\}  \nonumber \\ & \quad \quad \times \sum_{(\gamma',\gamma''|\gamma)} p (\gamma',\gamma''|\gamma) \bar \Pi(\gamma',t_i) \bar \Pi(\gamma'',t_i) \nonumber\\
%& + \left\{ 2 \sigma \Delta t_i \right\} \left\{ \bar \Pi(\gamma',t) \bar \Pi(\gamma'',t) \right\}, \nonumber \\
& + \sum_{e\in \mathcal{E}_i}\left\{\frac{\tau \Delta t_i }{N-1} \right\}  \sum_{(\gamma',\gamma''|\gamma)} p (\gamma',\gamma''|\gamma) \bar \Pi(\gamma',t_i) \Pi_{e}(\gamma'',t_i) \nonumber \\ 
& + \sum_{e\in \mathcal{E}_i} \left\{\frac{\tau \Delta t_i }{N-1} \right\}  \sum_{(\gamma',\gamma''|\gamma)} p (\gamma',\gamma''|\gamma) \Pi_{e}(\gamma',t_i) \bar \Pi(\gamma'',t_i) \nonumber \\
& + \sum_{e\in \mathcal{E}_i} \left\{\frac{\tau \Delta t_i }{N-1} \right\}\bar E(t_i) \Pi_e(\gamma,t_i ) \nonumber
\end{align}
where $\mathcal{E}_i(S) $ denotes the set of branches of $S$ in time slice $i$. The terms correspond to i) no event with an observed descendent; ii) birth of two gene lineages by speciation, duplication or transfer, such that both leave observed descendants; iii) and iv) birth of two gene lineages with observed descendants as a result of transfer back to the represented phylogeny; and finally, v)  transfer back to the represented phylogeny following which the copy in the unrepresented donor lineage does not leave an observed descendant, cf.\ Eq.5 and  Fig.\ A1 of \cite{szollosi_LGTftD}.

At speciation times $t=t_i$ where branches $f$ and $g$ descend from $e$ in $S$, a represented speciation takes place that may be followed by a loss, cf.\ Eq.6 and  Fig.\ A1 of \cite{szollosi_LGTftD}:
\begin{align}
\Pi_e(\gamma,t) &= \sum_{(\gamma',\gamma''|\gamma)} p (\gamma',\gamma''|\gamma)\Pi_f(\gamma',t) \Pi_g(\gamma'',t) \label{PieS}\\ & + \sum_{(\gamma',\gamma''|\gamma)} p (\gamma',\gamma''|\gamma)\Pi_f(\gamma'',t) \Pi_g(\gamma',t) \nonumber\\
 &+ \Pi_f(\gamma,t) E_g(t) + E_f(t) \Pi_g(\gamma,t).\nonumber 
\end{align} 

Finally at time $t=0$ on each terminal branch $e$ of $S$ the presence of observed genes for trivial clades $\gamma = \{ u\}$ comprised of a single leaf is expressed as:
\begin{align}
 \Pi_e(\{ u\},0) = 
\left\{
 \begin{array}{l l}
 1 & \text{if $u$ is a leaf of $G$ found in $e$ }\\
 0 & \text{otherwise} \\
 \end{array} \right. \label{Pie0}
\end{align}

\subsection{ALE implementation}

We implemented two methods to explore reconciliations for gene trees that can be amalgamated from an MCMC sample. Both of these methods take as their input a dated binary species tree and a set of conditional clade probabilities obtained from an MCMC sample of gene tree topologies. In both methods we set $\sigma=2N$, corresponding to making the assumption that the height of the species tree is equal to its expected value under the coalescent \cite{szollosi_LGTftD}.  Both implementations are in C++ and rely heavily on the  Bio++ library \cite{Dutheil:2006uq}. 

The first, which we call ALEsample samples duplication, transfer and loss rates using a simple Metropolis-Hastings algorithm \cite{metropolis:1087} using the likelihood $\mathcal{L}_{\mathrm{joint}} (A | S,\delta,\tau,\lambda,\sigma=2N) $ with an implicit flat prior on rates. At each step of the algorithm proposals are generated from the current rate values by adding a small random value to each of the three rates, boundaries at $0$ are considered as absorbing, i.e., for negative proposals a new proposal is generated. For a given set of DTL rates reconciled trees are sampled using stochastic backtracking along the dynamic programming sum \cite{szollosi_LGTftD}.       

The second, which we call ALEml optimises DTL rates using the downhill simplex method implemented in Bio++ by maximising $\mathcal{L}_{\mathrm{joint}} (A | S,\delta,\tau,\lambda,\sigma=2N)$ and subsequently finds the maximum likelihood (ML) reconciled gene tree for the ML set of rates using backtracking along the dynamic programming sum \cite{Szollosi:2012fkkk}. 

Our implementation of ALE is available from https://github.com/ssolo/ALE.git .

\section*{Maximum entropy distribution for marginal split frequencies}
We demonstrate that given marginal split frequencies the distribution over the space of all trees computed using conditional clade probabilities is the maximum entropy distribution. 

Consider $\mathcal{G}$ the set of all rooted trees with $n$ leaves, and denote by $N_{\mathcal{G}}$ the number of such trees. We index trees by $i=1\dots N_{\mathcal{G}}$. The indicator functions $\delta_i^\gamma=1$ and $\delta_i^\xi=1$ indicate, respectively, the presence of clade $\gamma$, and  the presence of split $\xi=(\gamma',\gamma''|\gamma)$ of clade $\gamma$ into complementary daughter clades $\gamma'$ and $\gamma''$, such that $\gamma \setminus \gamma'  =  \gamma''$ in tree $i$, and are $0$ for all other trees. To simplify notation we denote the sum over all splits of $\gamma$ as
\begin{equation}
\sum_{\xi \subset \gamma} \cdots  = \sum_{(\gamma',\gamma''|\gamma)} \cdots,\nonumber
\end{equation}
the sum over all possible splits as 
\begin{equation}
\sum_{\gamma,\xi} \cdots  = \sum_{\gamma\subset \Gamma} \sum_{\xi  \subset \gamma } \cdots,\nonumber
\end{equation}
and the set of splits in tree $i$ as $\xi \subset i$. Finally, we use the convention that identical lower-upper tree indices imply summation over all trees, e.g., $\delta_i^\xi p^i \equiv\sum_i \delta_i^\xi p^i$ and $ \delta_i^\gamma p^i\equiv\sum_i \delta_i^\gamma p^i$.  
 
Given an arbitrary probability distribution $p=\{ p^i \}$ on $\mathcal{G}$ the conditional clade probabilities are defined as 
\begin{align}
p_\xi &=\frac{  \delta_i^\xi p^i}{ \delta_i^\gamma p^i}.\label{pxi}
\end{align}

To derive the maximum entropy distribution given a set of observed marginal split frequencies $F_\xi $ we have to find among all distributions $p$ the distribution that  matches the observed split frequencies
and maximizes the entropy
\begin{equation}
- p_i\ln p^i \label{plnp}. 
\end{equation}
The entropy has to be maximised under the constraints of total probability, i.e. $\sum_i p^i=1$ and the observed split frequencies:
\begin{equation}
\delta_i^\xi p^i = F_\xi. \label{cF}
\end{equation}
 To find the maximum given the above linear constraints we maximize the Lagrangian
\begin{align}
&L =  - p_i\ln p^i -\alpha \left(\sum_i p^i - 1\right) - \sum_{\gamma,\xi} \lambda_\xi \left(  \delta_i^\xi p^i - F_\xi   \right). \nonumber
\end{align} 
Equating the derivative with respect to $p^i$ with zero gives:
\begin{align}
p^i &\propto \exp(-\sum_{\xi \subset i} \delta_i^\xi \lambda_\xi ) \propto \prod_{\xi \subset i} \Phi_\xi,\nonumber
\end{align}
where we define the notation $\Phi=\mathrm{e}^{-\lambda}$. Normalising $\Phi$-s such that 
\begin{equation}
\sum_{\xi  \subset \gamma } \Phi_\xi = 1  \label{phinorm}
\end{equation}
satisfies total probability. Furthermore, it implies that 
\begin{equation}
p_\xi = \Phi_\xi.\nonumber 
\end{equation}

To see that this is the case, we must consider that the branch at the base
of $\gamma$ defines an outer tree and an inner tree, the latter of which corresponds to the clade $\gamma$.  For any tree $i$ containing split $\xi$ of $\gamma$, and consequently also clade $\gamma$, one can write down products of $\Phi$-s such that all factors $\pi^i_\mathrm{out}$ corresponding to the outer tree are on the left, and all factors $\Phi_\xi \pi^i_\mathrm{in}$ corresponding to the inner tree are on the right, i.e., 
\begin{align}
p^i& = \pi^i_\mathrm{out} \Phi_\xi \pi^i_\mathrm{in}. \nonumber
\end{align}
For a given $\xi$ the outer tree is constrained only by the presence of $\gamma$ where as the inner tree is constrained by the presence of $\xi$. 
We now calculate the numerator and denominator of equation \ref{pxi} separately: 
\begin{align}
 &\delta_i^\xi p^i=  \delta_i^\xi \left( \pi^i_\mathrm{out}\Phi_\xi  \pi^i_\mathrm{in} \right), \nonumber\\
 &\delta_i^\gamma p^i= \delta_i^\gamma \left( \sum_{\xi' \subset \gamma}  \pi^i_\mathrm{out} \Phi_{\xi'} \pi^i_\mathrm{in} \right).  \nonumber
\end{align}
The sums $\sum_{i} \delta_i^\xi \cdots $ and  $\sum_{i} \delta_i^\gamma \cdots $ can both be split into an outer and an inner part such that the outer sum is over all trees that contain the clade $\gamma$  while the inner sum contains a particular split resolving $\gamma$:
\begin{align}
 &\delta_i^\xi p^i=  \delta_i^\gamma \pi^i_\mathrm{out} \Phi_\xi  \left(  \delta_j^\xi \pi^j_\mathrm{in} \right), \nonumber\\
 &\delta_i^\gamma p^i=  \delta_i^\gamma \pi^i_\mathrm{out} \sum_{\xi' \subset \gamma} \Phi_{\xi'} \left( \delta_j^{\xi'} \pi^j_\mathrm{in} \right) \nonumber
\end{align}
The inner sums can be calculated recursively starting from clades with only a single split for which $\Phi(\xi)=1$, for each ancestral  clade the normalisation in Eq.\ref{phinorm} recursively implies that these also sum to unity.   
It follows that  
\begin{align}
p_\xi = \frac{\delta_i^\xi p^i}{\delta_i^\gamma p^i } =\frac{\delta_i^\gamma \pi^i_\mathrm{out} \Phi_\xi}{\delta_i^\gamma \pi^i_\mathrm{out} }= \Phi_\xi.
\end{align}

\end{document}